\documentclass[letterpaper,12pt]{article}
\usepackage{amsmath}
\usepackage{amsfonts}
\usepackage{amssymb}
\usepackage{graphicx}
\usepackage{physics}
\usepackage{latexsym}
\usepackage{epsfig}
\usepackage{pstricks}
\usepackage{stmaryrd}
\usepackage{rotating}
\usepackage[english]{babel}
\usepackage{setspace}
\usepackage[utf8]{inputenc}
\usepackage{natbib}
\usepackage[T1]{fontenc}
\usepackage{lmodern}
\usepackage{enumitem}
\usepackage{csquotes}
\usepackage{amsthm}
\usepackage{xcolor}
\usepackage[margin=1in]{geometry}
\begin{document}
\begin{center}	
\begin{Large}
\textbf{A disputable assumption behind the empirical equivalence between pilot-wave theory and standard quantum mechanics}\\
\end{Large}
\end{center}

\begin{center}
\begin{large}
J. Manero, R. Muciño and E. Okon\\
\end{large}
\textit{Universidad Nacional Aut\'onoma de M\'exico, Mexico City, Mexico.}\\[1cm]
\end{center}

The de Broglie-Bohm pilot-wave theory asserts that a complete characterization of an $N$-particle system is given by its wave function \emph{together} with the (at-all-times-defined) positions of the particles, with the wave function always satisfying the Schrödinger equation and the positions evolving according to the deterministic ``guiding equation''. A complete agreement with the predictive apparatus of standard quantum mechanics, including the uncertainty principle and the probabilistic Born rule, is then said to emerge from these equations, without having to confer any special status to measurements or observers. Two key elements behind the proof of this complete agreement are \emph{absolute uncertainty} and the \emph{POVM theorem}. The former involves an alleged ``naturally emerging, irreducible limitation on the possibility of obtaining knowledge within pilot-wave theory'' and the latter establishes that the outcome distributions of all measurements are described by POVMs. Here, we argue that the derivations of absolute uncertainty and the POVM theorem depend upon the questionable assumption that ``information is always configurationally grounded''. We explain in detail why the offered rationale behind such an assumption is deficient and explore the consequences of having to let go of it.

\section{Introduction}
\onehalfspacing

The de Broglie-Bohm pilot-wave theory stands as the simplest and best-developed hidden-variable alternative to the standard formulation of quantum mechanics. It postulates that a complete and objective description of an $N$-particle system consists of two elements: the wave function and the actual positions of the particles---the latter assumed to exist at all times, even when not measured. The wave function is taken to always satisfy the Schrödinger equation and the positions to evolve according to the so-called guiding equation, a deterministic law fixing the velocity of each particle in terms of the wave function and the (instantaneous) positions of all particles.

In spite of being fully deterministic, the de Broglie-Bohm theory is said to fully reproduce the empirical predictions of standard quantum mechanics, including those typically associated with the uncertainty principle and the probabilistic nature of measurement outcomes described by the Born rule. Moreover, this agreement is argued to emerge, without having to assign any privileged role to the vaguely defined notions of measurement or observer. In other words, pilot-wave theory is claimed to fully recover the statistical and operational features of quantum mechanics from a fundamentally deterministic and observer-independent reality.

Two essential components of the proof of this purported empirical equivalence are the concept of \emph{absolute uncertainty} \citep{DGZ92} and the so-called \emph{POVM theorem} \citep{DGZ04}. The former is described as an inherent and unavoidable restriction on the ability of any observer to gain knowledge about a system's full state. Specifically, absolute uncertainty implies that there are fundamental constraints on empirical access to particle positions, even if this ``hidden variables'' are well-defined at all times. The notion of absolute uncertainty is then essential to justifying why, despite the deterministic evolution of the system's hidden variables (the particle positions), the observed behavior remains consistent with the standard probabilistic predictions of quantum mechanics. The POVM theorem is claimed to establish that the predictions of pilot-wave theory for all possible experiments are always describable in terms of positive operator-valued measures or POVMs and implies a strict condition for measurability. If that is the case, pilot-wave would be able to derive or explain the essential role played by operators in the predictive apparatus of standard quantum mechanics.

However, here we argue that the derivations of absolute uncertainty and the POVM theorem crucially rely on a questionable assumption, namely, the claim that ``information is always configurationally grounded''---meaning, in particular, that the most complete knowledge an external observer could have about a system is necessarily contained (or exhausted) by the actual particle configuration of the system's environment. We provide a detailed analysis of why the rationale behind such an assumption is flawed, challenging the general validity of the proofs. Furthermore, we explore the profound implications of questioning these key results, particularly in relation to the extent to which the pilot-wave framework really coincides with the standard quantum mechanical predictive apparatus.

Our manuscript is organized as follows. In section \ref{PW}, we present an overview of the pilot-wave derivations of absolute uncertainty and the POVM theorem. Then, in order to lay the groundwork for a careful analysis of such derivations, in section \ref{Det} we make some general remarks about the notion of detection. Next, in section \ref{UA}, we expose what we take to be a questionable assumption in the proofs and, in section \ref{Cons}, we explore some consequences of challenging absolute uncertainty and the POVM theorem. Finally, section \ref{Conc} contains our conclusions.

\section{Pilot-wave theory, absolute uncertainty and the POVM theorem}
\label{PW}

According to the de Broglie-Bohm pilot-wave theory, the complete description of a non-relativistic $N$-particle system is given by its wave function, $ \Psi (q, t) $, together with the positions of the particles, $Q(t)=\left(\mathbf{Q}_1 (t),\mathbf{Q}_2 (t),\dots,\mathbf{Q}_N (t)\right)$---taken at all times to possess well-defined values. The wave function is postulated to always satisfy the Schr\"odinger equation
 \begin{equation}
 \label{Sch}
 i \hbar \frac{\partial \Psi}{ \partial t } = - \sum_{k=1}^N \frac{\hbar^2}{2 m_k} \nabla^2_k \Psi + V(q)\Psi 
 \end{equation}
and the positions to evolve according to the guiding equation, 
\begin{equation}
\frac{d \mathbf{Q}_k}{dt} = \frac{\hbar}{m_k} \left. \text{Im}\left[ \frac{\nabla_k \Psi}{ \Psi} \right] \right\rvert_{q=Q(t)}.
\label{guide}
\end{equation}
Finally, the theory assumes, at some initial time $t=0$, the distribution of the positions of the particles to satisfy 
$\rho_0(q) =|\Psi_0(q)|^2$. This assumption, together with the guiding equation and the quantum continuity equation, implies so-called \emph{quantum equilibrium}; namely, that at all times it is the case that the distribution of particles satisfies $\rho_t(q) =|\Psi_t(q)|^2$. 

It seems clear that quantum equilibrium takes us some way towards an empirical agreement between pilot-wave theory and the standard quantum formalism. However, many details remain to be filled. Two important results are argued to complete the job: absolute uncertainty \citep{DGZ92} and the POVM theorem \citep{DGZ04}. Next, we review them both.

\subsection{Absolute uncertainty}

Following \cite{DGZ92}, we start by noting that any subset of an $N$-particle system induces a splitting
\begin{equation}
 q=(x,y) ,
\end{equation}
with $x$ the generic configuration of the particles of the subsystem and $y$ that of its complement or environment. The splitting also induces a splitting
\begin{equation}
 Q=(X,Y) ,
\end{equation}
with $X$ and $Y$ the actual configurations of the subsystem and its environment.

Consider, now, what can be said about the distribution of the particles of the subsystem (assuming knowledge of the total wave function $\Psi_t(q)$). Given quantum equilibrium, we know that, at all times, the distribution of the whole system satisfies $\rho_t(q) =|\Psi_t(q)|^2$. It clearly follows that 
\begin{equation}
\mathbf{P}(X_t \in dx | Y_t) = |\Psi_t(x,Y_t)|^2 dx.
\end{equation}
Defining the conditional wave function by $\psi_t(x) \equiv \Psi_t(x,Y_t)$, it then follows that 
\begin{equation}
\label{FPF}
\mathbf{P}(X_t \in dx | Y_t) = \mathbf{P}(X_t \in dx | \psi_t(x))   = |\psi_t(x)|^2 dx .
\end{equation} 

Equation (\ref{FPF}) is called the \emph{fundamental conditional probability formula}, and is said to be the cornerstone of the analysis in \cite{DGZ92} regarding the emergence of randomness in pilot-wave theory. In particular, it implies that  $X_t$ and $Y_t$ are conditionally independent, given the conditional wave function $\psi_t(x)$.

The next step in \cite{DGZ92} is to explore the thorny notion of \emph{knowledge} and to argue that, regardless of how one cashes out the idea of (external) knowledge about the state of a given system, such knowledge must be fully contained in the \emph{configuration} of its environment. That is, it is argued that, whatever one may mean by knowledge of a system, it must be grounded or encoded in the configuration of things external to it.

By combining Eq. (\ref{FPF}) and the provision that all knowledge of a system must be grounded in the configuration of its environment, one arrives at \emph{absolute uncertainty}, which contends that quantum equilibrium conveys the most detailed knowledge possible concerning the present configuration of a subsystem. That is, that any information additional to $\psi_t(x)$ one might have can be of no relevance whatsoever to the possible value of $X_t$. In \cite{DGZ92}, the result is stated as follows:
\begin{quotation}
We are thus claiming to have established that in a universe governed by Bohmian mechanics it is in principle impossible to know more about the configuration of any subsystem than what is expressed by $\rho=|\psi|^2$---despite the fact that for Bohmian mechanics the actual configuration is an objective property, beyond the wave function.
\end{quotation}

\subsection{The POVM theorem}
\label{POVM}

Arising from an analysis of the notion of ``measurement'' in the context of pilot-wave theory, the POVM theorem states that the predicted statistics for an experiment are always described by a POVM acting on the Hilbert space of the measured system (the analysis was first outlined in \cite{Bohm} and developed in detail in \cite{DGZ04}; see \cite{Dustin} for a recent thorough assessment of the theorem and \cite{Lazarovici2020} for a clear exposition of the standard view regarding the epistemic status of positions in pilot-wave theory).
 
In broad terms, the idea of the analysis is the following. A measurement scenario is characterized as an interaction between some (typically microscopic) system and a macroscopic measuring apparatus. The system+apparatus duo is assumed, to a very good approximation, to be isolated from the rest of the world (with initial wave function $\Psi_0(x,y)$, with $x$ and $y$ the configurations of  system and apparatus, respectively). Finally,  the initial conditional wave function of the system is assumed to be an effective wave function $\psi(x)$.\footnote{A subsystem (with generic configuration $x$) is said to possess an effective wave function $\phi(x)$, when the total wave function can be written as
\begin{equation}
\Psi(x,y) = \phi(x) \Phi(y) + \Psi^{\perp}(x,y)
\end{equation}
with $\Phi$ and $\Psi^{\perp}$ having macroscopically disjoint $y$-supports, and $Y$ in the support of $\Phi$ (see \cite{DGZ92} for details).
} 

Then, the system and apparatus interact, and the result of the measurement is assumed to end up encoded in the final \emph{configuration} of the apparatus. Quoting \cite{DGZ04}:
\begin{quotation}
It is important to bear in mind that regardless of which observable one chooses to measure, the result of the measurement can be assumed to be given configurationally, say by some pointer orientation or by a pattern of ink marks on a piece of paper.
\end{quotation}

In more detail, as a result of the interaction, the system+apparatus state evolves into an entangled superposition
\begin{equation}
\Psi_0(x,y) \rightarrow \Psi_T(x,y) = \sum_k \alpha_k \psi_k (x,y)
\end{equation}
with the $\psi_k(x,y)$  localized in macroscopically disjoint regions of the apparatus configuration space. Then, applying quantum equilibrium, the outcome---the final pointer orientation---is effectively random, governed by $|\Psi_T|^2$.  As a result, the probabilities for the final position of the pointer are given by $p_k = \int_{G_{k}} |\Psi_T(x,y)|^2 dx dy$ with $G_k$ the disjoint regions in the apparatus configuration space, corresponding to the different $y$-supports of the $\psi_k(x,y)$.

Summing up, performing an experiment on a system with initial effective wave function $\psi$ leads to the result $Z$ encoded in the final configuration $Q_T$, i.e., $Z = F (Q_T)$. Since $Q_T$ is randomly distributed according to the quantum equilibrium measure $\rho_T$, the probability distribution of $Z$ is given by the induced measure $\rho_\psi^Z = \rho_T \circ F^{-1}$. Now, since the map $\psi \rightarrow \rho_\psi^Z$ is quadratic in the initial wave function $\psi$, by the Riesz representation theorem it is equivalent to a POVM. Therefore, the predictions of pilot-wave theory for any experiment are described in terms of some POVM.

An important consequence of the POVM theorem \citep[section 7.1]{DGZ92} is that it can be used to construct a necessary condition for \emph{measurability}:
\begin{quotation}
A quantity is not measurable when there is a value for it, which is possible when the wave function is $\psi_1 + \psi_2$, but not  possible when the wave function is either $\psi_1$ or $\psi_2$.
\end{quotation}
In particular, since every wave function $\psi$ may be written as a sum of its real and imaginary parts, and the guiding equation imposes zero velocity for all real or imaginary wave functions, the condition implies that velocity is not measurable.

\section{Preparatory remarks about detections}
\label{Det}

In order to lay the groundwork for a dissection of the pilot-wave derivations of absolute uncertainty and the POVM theorem, we make some general comments about the notion of \emph{detection} (or \emph{measurement}).

Broadly speaking, we say that a device M \emph{detects} a feature of a system S when, through their interaction, certain properties of S---which may not have existed prior to the interaction---become correlated with certain properties of M. When this happens, those correlated properties of M are said to codify or encode the outcome of the detection. For example, when an ammeter measures electric current in a circuit, the needle's position on the dial of the ammeter becomes linked to the current's value, thereby encoding it. Likewise, when someone sees a black cat crossing the street, that observation is encoded in their brain's physical state (in this broad framework, human perception is also a form of detection, with the brain acting as the detecting and encoding system). Naturally, a thorough account of detection would need to address issues like observability, control, consistency, and the reliability of the correlation. But for our current discussion, this general characterization is sufficient (see \cite{Roberts, Greaves, Dasgupta2016, Teh2016, Wallace, MurgueitioRamirez2022, Baker2023WhatAS, Sebas} for an engaging, contemporary discussion regarding the relation between symmetries and measurable quantities).

The description of detection given above highlights that it is the \emph{dynamics} of a theory that dictates which devices can measure which properties. For a property of system S to be detectable by device M, the necessary kind of interaction between S and M must be allowed by the theory. In addition, the dynamics determines which specific aspects or degrees of freedom of M are capable of encoding the outcome of the detection.

With this in mind, imagine a world governed by a specific theory that features a set of degrees of freedom, which can be naturally grouped into different categories (e.g., position and velocity, particle and field, material and gravitational, fields A, B, and C, etc.). Now, suppose one wishes to assert that all measurement outcomes---or more broadly, all information---can be expressed using only one particular type of degree of freedom (for instance, solely in terms of positions, particles, field A, etc.). For this assumption to hold, the theory's dynamics must support the formation of the necessary correlations between the relevant quantities, enabling the chosen type of degree of freedom to effectively encode all information.

Consider a scenario where system S interacts with device M. If one wishes to claim that the outcome of any measurement performed by M on S can be captured using, for example, the $r$-type degrees of freedom of M, then it must be demonstrated that an interaction can be established in which every objective property of S becomes correlated with those $r$-type degrees of freedom of M. If such a correlation cannot be achieved for some property of S---that is, if the dynamics of the theory does not allow it---then one cannot conclude that such property itself is fundamentally undetectable. Rather, the proper conclusion would be that the assumption that all measurement results can be expressed solely in terms of the $r$-type degrees of freedom is unfounded. Let's now consider some specific examples to illustrate this point.

Imagine a world containing a particle with two distinct types of properties, A and B, where there are no inter-type interactions. Now, suppose we arbitrarily require that all measurement outcomes must be expressed in terms of, say, A properties. Given the lack of interaction between A and B properties, it would then appear that B properties are undetectable. However, this apparent undetectability does not stem from any fundamental unobservability or lack of objectivity of B's properties---it simply results from the imposed constraint to use only A-type properties to record information. Since the theory postulates A and B properties as equally real and objective, there is no legitimate basis for privileging one over the other in terms of detectability.

More concretely, consider a model containing particles with position degrees of freedom, as well as some non-configurational degrees of freedom (e.g., velocities or some other internal properties). Suppose, in addition, that the dynamics of the model is such that the positions of different particles can get correlated, and the same for the non-configurational degrees of freedom, but that the position of one particle does not interact with the non-configurational degree of freedom of another. Finally, suppose that it is asserted that results of all measurements are to be recorded in terms of positions only. Under such circumstances, it is clear that the non-configurational degrees of freedom would be undetectable. However, as with the example above, this undetectability would only arise as a consequence of the arbitrary decision to record all results in terms of positions. If results are allowed to be encoded in terms of positions \emph{and} non-configurational degrees of freedom, both perfectly real and objective properties, then everything would be detectable.

Finally, consider a Newtonian theory of point particles. In that case, the degrees of freedom can be separated into positions and velocities. Suppose one wants to impose that all measurements can be recorded purely in terms of positions; is such an assumption justified? Above we considered examples in which this sort of restriction is not valid, but this case is different.  As we saw, in order to evaluate the validity of an assumption like this, one has to explore the dynamics and, in this Newtonian case, the dynamics seems generically capable of providing correlations between the positions \emph{and} velocities of some subset of particles (the system measured), with only positions of a different, larger, subset of particles (the measurement device). In particular, the dynamics seems to allow for the construction of speedometers, in which the velocity of a system (e.g., a car) is recorded in terms of the position of another system (e.g., a needle in a dial). It seem, then, that, in this case of a Newtonian theory of point particles, measurements of all properties can indeed be recorded configurationally.

Does this mean we can always adopt the assumption that outcomes of measurements can always be recorded in configurational terms? Certainly not. As was made clear above, the validity of any assumption that restricts the degrees of freedom available for registering measurement outcomes depends on the specific dynamics of the theory under consideration. And, for any given theory, the justification for imposing limitations on the types of admissible records cannot rely on what holds in the Newtonian framework previously discussed, nor on empirical facts about our actual world. In other words, one must not conflate the theoretical question of which degrees of freedom a given dynamics allows to become correlated, with the empirical matter of what can be observed in our physical reality. Furthermore, if the predictions of a framework regarding observable records fail to align with empirical observations, the conclusion should not be that arbitrary restrictions on record types are permissible, but rather that the model in question lacks empirical adequacy. With this clarified, we are now ready to examine the specific case of pilot-wave theory.

\section{A questionable assumption behind absolute uncertainty and the POVM theorem}
\label{UA}

In this section, we expose a key assumption behind the pilot-wave derivations of absolute uncertainty and the POVM theorem, and we assess whether such an assumption is justified.\footnote{The POVM Theorem has also been under fire recently due to the ``Arrival-times'' controversy; see \cite{DasDurr, Gold1, Gold2, Das, sym16101325}.} 

\subsection{The configurations-only assumption}\label{ssection:configurationsassumption}

As we saw above, absolute uncertainty can be understood as following from two premises:
\begin{enumerate}
\item The fundamental conditional probability formula 
\begin{equation}
\mathbf{P}(X_t \in dx | Y_t)  = |\psi_t(x)|^2 dx.  \nonumber
\end{equation} 
\item The assumption that ``information is always configurationally grounded''.
\end{enumerate}
We already saw that the first premise is always true within pilot-wave theory. That is, conditional on $Y_t$, it is always the case that what can be said about the distribution of the particles of a system is exhausted by the Born rule. The second premise, on the other hand, is less safe, as it goes beyond first principles of pilot-wave theory. It asserts that all knowledge (or information) about the state of the system that external agents could in principle gather, must be contained in the \emph{positions} of the particles of the environment of the system. It is only because of this constraint on how information can be encoded that the innocent looking formula above turns into an in principle limitation on the possibility of knowing about something postulated to be real and objective within the theory.

If, in contrast, one allows for information to be grounded in non-configurational features of the environment---in particular, in \emph{velocity} patterns---then one easily escapes the knowledge limitation. This is because, in general, $\dot{Y}_t$ is a function of $X_t$. Therefore, knowledge about $\dot{Y}_t$ may contain information about $X_t$.

As a simple example, consider two particles in one dimension (with positions $x$ and $y$ respectively), in the state
\begin{equation}\label{examplestate}
\Psi(x,y) = \frac{1}{\sqrt{2}} \left[\phi_a(x) \varphi_p(y) + \phi_{-a}(x) \varphi_{-p}(y) \right]
\end{equation} 
with $\phi_{\pm a}$ a wave function with highly localized position around $\pm a$ and $\varphi_{\pm p}$ a wave function with highly localized momentum around $\pm p$. In this case, compatible with the fundamental conditional probability formula, knowledge about $Y$ would not provide any information regarding $X$ beyond what can be inferred from the state above---namely, a 50/50 chance of $\pm a$. However, knowledge of $\dot{Y}$ would immediately allow to determine whether $a$ or $- a$ is the case, breaking the limitation imposed by absolute uncertainty.

At this point, it could be argued that, since pilot-wave is a first-order theory, in which the wave function and positions at time $t$ fully determine all velocities at time $t$, knowledge of $Y$ already contains all information about $\dot{Y}$. Note however that, in the presence of entanglement between the system and its environment, the fundamental non-locality of the theory makes $X$ necessary to compute $\dot{Y}$. Therefore, in general, knowledge of $Y$ does not allow to infer $\dot{Y}$. From all this, it follows that, in the absence of entanglement between a system and its environment, it is true that all an external agent can say about the state of a system in exhausted by the Born rule. However, that is not true in general: information about positions and velocities of an environment does not limit knowledge of the positions of a system to quantum equilibrium.

Now, regarding the POVM theorem, the starting point of the analysis is the idea that a quantum measurement is an interaction between a fairly well-isolated system+apparatus pair, such that the result of the measurement is read from the final configuration or ``pointer position'' of the apparatus. Since the pair is assumed to be well-isolated, one can impose quantum equilibrium on the final distribution of positions, so the outcomes can be predicted via the Born rule applied to the pointer positions of the apparatus. 

However, it is only because it is assumed that the result is encoded by the final \emph{position} of the pointer, and not other property, such as its velocity, that quantum equilibrium becomes relevant and one arrives at the quadratic map $\psi \rightarrow \rho_\psi^Z$ and the POVM theorem. In contrast, if the result is allowed to be encoded in terms of the velocity of the pointer, then the limitation imposed by absolute uncertainty would not apply and the map from the initial wave function to the final prediction need not be quadratic, blocking the derivation of the POVM theorem.

At this point, one could argue that, even if at some time, a result is recorded in terms of, both, the position and velocity of a pointer, it is always possible to then record that information in terms of only positions of a different pointer. If so, the assumption that all results can be recorded in terms of positions would be justified.  That is, suppose that system S (with coordinates $x$) interacts with measuring apparatus M (with coordinates $y$) and that at time $T_1$, after the interaction, the result of the measurement is recorded in terms of, both, $Y$ and $\dot{Y}$. Is it then possible, in general, to make M interact with some other apparatus, N (with coordinates $z$), such that $Y_{T_1}$ and $\dot{Y}_{T_1}$ end up encoded in terms of the values of $Z$ at some later time $T_2$. If that would be possible, then the configurations-only assumption would be justified, as it would always be possible to encode results in terms of configurations only. That is, the interaction between S and M above would only represent an internal, intermediate step, but N would be the true measuring apparatus all along---with results encoded in purely configurational terms, as demanded.

However, employing the resources of \cite{DGZ04}, it can be shown that, in general, $Y_{T_1}$ and $\dot{Y}_{T_1}$ cannot be recorded in terms of $Z_{T_2}$. To see it, we recall that, with the configurations-only assumption in place, the measurability condition derived from the POVM theorem establishes that velocity cannot be encoded in terms of positions. This implies, in particular, that $\dot{Y}_{T_1}$ cannot be generically recorded in terms of $Z_{T_2}$. We see that, although the POVM is not true in general, because it depends upon an unjustified assumption, the results in \cite{DGZ04} can be leveraged to show that velocity patterns cannot be translated into configuration patterns. 

We conclude that, both, absolute uncertainty and the POVM theorem crucially depend upon the assumption that all information must always be configurationally grounded---i.e., that the most complete knowledge an external observer could have about a system is necessarily contained by the actual particle configuration of the system's environment. The key question we must turn to now, then, is whether this configurations-only assumption is justified.

\subsection{Is the assumption justified?}

The configurations-only assumption demands for the results of all possible experiments to be recorded in purely configurational terms. The discussion in section \ref{Det} makes it clear that, in order for an assumption of that sort to be justified, one has to look at the \emph{dynamics} of the theory in question. In particular, in order to justify the configurations-only assumption, one must check whether the dynamics of pilot-wave theory allows for measuring scenarios to be established, such that all (not-necessarily pre-existing) real, objective properties of a system can end up correlated with the position of some pointer. If this is not the case---as with the simple examples in section \ref{Det}---one cannot conclude that the properties that do not end up correlated with positions are undetectable. Instead, one must conclude that the configurations-only assumption is not justified. Therefore, it is not that absolute uncertainty and the POVM theorem establish deep, in principle limitations on the possibility of obtaining knowledge; the limitation arises from the arbitrary restriction to record everything in terms of positions.

It could be argued that the restriction is reasonable because it aligns with experience: we seem to do just fine by recording every sort of information in configurational terms, and it's not even clear what would it mean for information not to be recordable configurationally. However, as argued above, the rationale for restricting the kinds of admissible records within an hypothetical theory---pilot-wave theory in this case---cannot be based on empirical facts specific to our actual world. One must distinguish between the theoretical issue of which degrees of freedom a given dynamics permits to become correlated and the empirical issue of what can be observed in our world. Moreover, if a framework's predictions about observable records do not match real-world observations, this should not justify imposing \emph{ad hoc} limitations on the types of records considered. Instead, it indicates that the framework itself might not be empirically adequate.

Moreover, it is not really the case that, in practice, we are only capable of observing positions of things; we also seem capable of directly observing their velocities, among other things. That is, it is not as if, empirically, velocities are hidden from us. Therefore, even going by what we can observe in practice, the configurations-only assumption seems unwarranted. More to the point, for all we know, perceptions might not supervene on instantaneous brain states, so it is not clear whether information in human brains is really encoded in purely configurational terms.

Having said all this, it is true that it would be a prediction of pilot-wave theory that \emph{there are things that are knowable, but not communicable in configurational terms}. This sounds like an extremely bizarre and alien situation, radically different from anything we have experienced. We stress, however, that this would be the predictions of pilot-wave theory; and if they do not align with experience, then it is the theory that bears the brunt.

\subsubsection{Justification by Dürr et al.}

We close this section by assessing the justification for the configurations-only assumption provided by \cite{DGZ92}, contained in footnotes 9 and 42. We quote both footnotes in full, intercalating them with our observations.

Footnote 9 starts as follows:
\begin{quotation}
\noindent [The absolute uncertainty] argument appears to leave open the possibility of disagreement when the outcome of the measurement is not configurationally grounded, i.e., when the apparatus variables which express this outcome are not functions of $q_{app}$.
\end{quotation}
It is clear that \cite{DGZ92} is well-aware that the configurations-only assumption is needed in order to prove absolute uncertainty. Nevertheless, it attempts to justify the assumption as follows:
\begin{quotation}
\noindent However, the reader should recall Bohr’s insistence that the outcome of a measurement be describable in classical terms,...
\end{quotation}
We see the appeal of using Bohr's own ideas to argue against standard quantum mechanics. However, we doubt one really wants the empirical equivalence between pilot-wave theory and standard quantum mechanics to depend upon such vague notions such as the quantum/classical divide. In any case, in a universe governed by pilot-wave theory, descriptions of all scenarios---including measurements---are never classical, they are always pilot-wave descriptions.

The footnote continues:
\begin{quotation}
\noindent ...as well as note that results of measurements must always be at least potentially grounded configurationally, in the sense that we can arrange that they be recorded in configurational terms without affecting the result.
\end{quotation}
This is a statement of the assumption, not a justification thereof. That is, it is no more than the stipulation that measurements ``must always'' possess certain features. In any case, we saw above that, as a result of the pilot-wave dynamics, measurements of velocities simply cannot be recorded in configurational terms. And, as argued in section \ref{Det}, this does not mean that velocities are not detectable, but that the assumption that all measurements can be grounded in configurations is unjustified.
 
The footnote ends with:
\begin{quotation}
\noindent Otherwise we could hardly regard the process leading to the original result as a completed measurement.
\end{quotation}
This seems to be circular: measurements must end with written results because, otherwise, one cannot regard measurements as completed.

Footnote 42 begins as follows:
\begin{quotation}
The reader concerned that we have overlooked the possibility that information may sometimes be grounded in non-configurational features of the environment, for example in velocity patterns, should consider the following...:

(1) Knowledge and information are, in fact, almost always, if not always, configurationally grounded. Examples are hardly necessary here, but we mention one---synaptic connections in the brain.
\end{quotation}
It is not hard to find examples in which information is not configurationally grounded, e.g., FM radio, in which information is encoded in terms of frequencies. Moreover, we find the brain example quite problematic---for all we know, perceptions might not supervene on instantaneous brain states (e.g., feedback loops might be essential).
\begin{quotation}
(2) Dynamically relevant differences between environments, e.g., velocity differences, which are not instantaneously correlated with configurational differences quickly generate them anyway. And we need not be concerned with differences which are not dynamically relevant!
\end{quotation}
As we saw above in detail, according to the pilot-wave dynamics not all velocity patterns translate into configurational patterns. 
\begin{quotation}
(3) Knowledge and information must be communicable if they are to be of any social relevance; their content must be stable under communication. But communication typically produces configurational representations, e.g., pressure patterns in sound waves.
\end{quotation}
Above we concluded that, in a pilot-wave universe, there would be things that are knowable but not communicable in configurational terms. And, that, surely sounds quite alien. However, as we explained in section \ref{Det}, one must not confuse the theoretical question about what would be observable in a pilot-wave universe, with empirical facts of our world. And, if a framework's predictions do not match what is actually observed, the appropriate conclusion is not that one can arbitrarily limit the types of records considered, but that the model itself fails to adequately reflect reality.
\begin{quotation}
(4) In any case,... when a system has an effective wave function, the configuration $Y$ provides an exhaustive description of the state of its environment...
\end{quotation}
Again, this is circular: it is exhaustive, only if ``information is always configurationally grounded''.

\section{Consequences}
\label{Cons}

Absolute uncertainty is said to impose an in principle limit on what can be known. Similarly, the POVM theorem is argued to severely restrict the set of properties that can be measured. Both results are essential to ``hide'' the extra structure within the pilot-wave framework and deliver an empirical agreement between the theory and standard quantum mechanics. We argued, however, that these results depend upon the assumption that information is always configurationally grounded---an assumption that cannot be justified from first principles. We grant that, with the configurations-only assumption in place, the predictions of pilot-wave theory do coincide with those of standard quantum mechanics, but when such key assumption is lifted, then all bets are off. Below, we briefly explore the potential consequences of letting go of absolute uncertainty and the POVM theorem.

\subsection{Uncertainty principle}

As explained in the previous sections, without the configurations-only assumption we can gather more information about the actual configuration $X_t$ of a subsystem than what is contained in $|\psi_t(x)|^2$. Measurements whose records are not strictly configurational, but include, for example, the velocities $\dot{Y}$ of the environment allow, for additional information about the positions $X_t$ than what is contained in $|\psi_t(x)|^2$ to be gathered. For instance, in a two-path experiment, this additional information makes it possible to know which path was actually taken by a given particle (see, for instance, the state (\ref{examplestate}) of the example in section \ref{ssection:configurationsassumption}). That is, in pilot-wave theory, the particle trajectories are not so hidden after all.

As a result, the following assertion in \cite{DGZ92} is not justified:
\begin{quotation}
...from the perspective of a Bohmian universe the uncertainty principle is sharp and clear. In particular, from such a perspective it makes no sense to try to devise \emph{thought} experiments by means of which the uncertainty principle can be evaded, since this principle is a mathematical consequence of Bohmian mechanics itself. One could, of course, imagine a universe governed by different laws, in which the uncertainty principle, and a great deal else, \emph{would} be violated, but there can be no universe governed by Bohmian mechanics---and in quantum equilibrium--- which fails to embody absolute uncertainty and the uncertainty principle which it entails.
\end{quotation}
According to our analysis, in contrast, in a Bohmian universe the uncertainty principle is not a mathematical truth; particle configurations can be in quantum equilibrium and, still, the acquisition of knowledge associated with the configuration of a subsystem may not be strictly mediated by its wave function, meaning that $\psi$ does not represent maximal information about its configuration.

\subsection{Signaling}

It is also commonly argued that, when quantum equilibrium holds, the absence of superluminal signaling is a necessary consequence of pilot-wave theory (see \cite{Valentini}). However, such a result crucially depends on absolute uncertainty so, without it, superluminal signaling can no longer be shown to be forbidden. To see this, it should be noted that, in all demonstrations arguing for the absence of superluminal signaling, it is (often implicitly) assumed that $\psi_t(x)$ conveys maximal information about the configurations $X$. Provided that the configurations-only assumption is not justified, the actual position of the particles could be measured to a better degree than what the Born rule dictates and, with this, a protocol for sending superluminal signals could be constructed.
 
\subsection{Preferred foliations}

In addition to the non-relativistic theory discussed above, there are relativistic pilot-wave models, where a wave function or functional, satisfying an appropriate relativistic wave equation, ``guides the motion'' of either particles or fields. The formulation of the corresponding guide equation requires something like a privileged foliation of spacetime to play the role that absolute simultaneity plays in the non-relativistic theory, a fact that seems to imply a violation of Lorentz invariance. It has been argued, though, that this preferred foliation is completely undetectable, so these models are taken to be fully Lorentz invariant at the \emph{empirical} level. The problem is that, once more, such an empirical undetectability crucially depends upon the configurations-only assumption.

In more detail, the issue is the following. It can be shown that if one assumes the density of crossings to satisfy the standard quantum statistics at an initial leaf of the foliation, $\Sigma_0$, then this will be so for all leafs of the foliation. This is not true, though, for all hypersurfaces. In fact, in \cite{berndl} it is shown that the quantum distribution cannot simultaneously be realized in all Lorentz frames. In spite of this, what can be shown is that, when \emph{position measurements} are performed along any hypersuface, the density of crossings measured always correspond to the standard quantum results. From this, together with the claim that the actual positions of the particles cannot be measured to a better degree than what $\psi_\Sigma(x)$ predicts, it is concluded that the foliation is undetectable.

That is, by stipulation, the wave function satisfies a relativistic wave equation, so it is Lorentz invariant.\footnote{This is always true of the universal wave function, but it is also true of the apparatus+system wave function in a measurement, since such a pair is assumed to be isolated from the environment.}  This, plus the claim---arising from absolute uncertainty and the POVM theorem---that the wave function fully mediates all predictions, implies the non-detectability of the foliation. If, in contrast, one can know about positions more than what the wave function implies, then there are no reasons to conclude that the preferred foliation would be empirically invisible. In particular, if one were to measure velocities of particles, the fact pointed out in \cite{berndl} that the Born rule is nor valid in every hypersurface, would be empirically accessible---opening the door for superluminal signaling.


\section{Conclusions}
\label{Conc}

Two key elements underpinning the proof of complete agreement between pilot-wave theory and standard quantum mechanics are the concept of absolute uncertainty and the POVM theorem. The principle of absolute uncertainty refers to a purportedly fundamental and irreducible limit to the knowledge one can obtain about the conditions of a system within the framework of pilot-wave theory. It is claimed that this limitation emerges naturally from the dynamics of the theory and serves to ensure compatibility with the statistical predictions of quantum mechanics. The POVM theorem, on the other hand, plays a central role in the pilot-wave formalism by demonstrating that the statistical distributions of outcomes for all conceivable quantum measurements are represented by POVMs. This theorem is invoked to argue for the claim that the measurement outcomes in pilot-wave theory match those predicted by standard quantum mechanics.

However, in this work, we argue that the derivations of both absolute uncertainty and the POVM theorem rest upon a questionable and largely unexamined foundational assumption: that information is always configurationally grounded---that is, that all relevant information about a quantum system must be encoded in the configuration (or position) of the particles in its environment. This assumption effectively restricts the representational basis of information to spatial configurations, excluding alternative forms of encoding it.

We provide a detailed critique of this assumption, highlighting conceptual and methodological shortcomings in the arguments typically used to justify it. In particular, we show that the rationale for configurational grounding often relies on circular reasoning or appeals to empirical evidence that do not withstand closer scrutiny. By challenging the necessity of configurational grounding, we open the door to reconsidering foundational aspects of pilot-wave theory and quantum measurement more broadly. We also explore the potential consequences of abandoning this assumption, which include potential superluminal signaling, detection of the preferred frame, as well as a reevaluation of the purported equivalence between pilot-wave theory and orthodox quantum mechanics.



\bibliographystyle{apalike}
\bibliography{bibEEPW.bib}

\end{document}